\documentstyle[12pt,psfig]{article}
\newcommand{\beq}{\begin{equation}}
\newcommand{\eeq}{\end{equation}}
\newcommand{\bearr}{\begin{eqnarray}}
\newcommand{\eearr}{\end{eqnarray}}
\newcommand{\BE}{\begin{eqnarray}}
\newcommand{\EE}{\end{eqnarray}}
\newcommand{\BES}{\begin{eqnarray*}}
\newcommand{\EES}{\end{eqnarray*}}

\newcommand{\erf}{\mbox{erf}}

\newcommand{\befig}[1]{\begin{figure}[#1]}
\newcommand{\eefig}{\end{figure}}

\newcommand{\LL}{\left}
\newcommand{\RR}{\right}
\newcommand{\BSP}{\begin{samepage}}
\newcommand{\ESP}{\end{samepage}}

\begin{document}
\begin{titlepage}
\title{Quantum Molecular Dynamics of Partially Ionized Plasmas} 
\author{W. Ebeling\thanks{e-mail:
    werner@summa.physik.hu-berlin.de}~~and B. Militzer\thanks{e-mail:
    militzer@summa.physik.hu-berlin.de}\\ 
      Institute of Physics \\ Humboldt University Berlin\\ 
        Invalidenstr. 110, D-10115 Berlin, Germany}
\date{January 24, 1997}
\end{titlepage} 
\maketitle
\newcommand{\leqs}{$\stackrel{\scriptstyle<}{\scriptscriptstyle\sim}\:$}
\begin{abstract}
  \noindent
  We study a partially ionized hydrogen plasma 
  by means of quantum molecular dynamics,
  which is based on wave packets. We introduce a new model which 
  distinguishes between free and bound electrons. The free electrons
  are modelled as Gaussian wave packets with fixed width. For the bound
  states the 1s-wave function of the hydrogen atom is assumed. In our 
  simulations we obtain thermodynamic properties in the equilibrium
  such as the internal energy and the degree of ionization. The
  degree of ionization is in good agreement with theoretical predictions.
  The thermodynamic functions agree well with results from quantum
  statistics for 10000K~\leqs T~\leqs 40000K.
\end{abstract}
\vspace*{1cm}
Key words:\\
\\
quantum molecular dynamics, wave packet simulation, dense plasma, 
thermodynamic properties of plasmas, degree of ionization, 
ionization and recombination processes\\
\newpage
\section{Introduction}
\begin{figure}[htb]
  \vspace*{-1.5cm}
  \centerline{\psfig{figure=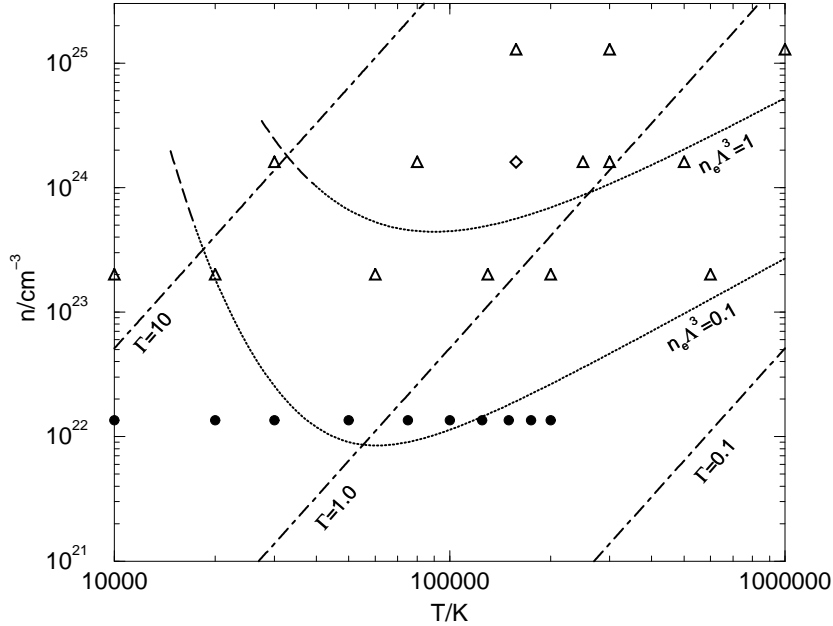,width=12cm,angle=270}}
  \caption{
   Characteristic lines of the hydrogen plasma in the temperature-density
   plane and data points of different investigations:
   $\bullet$ indicate our calculations, 
   $\scriptstyle\bigtriangleup$ are those of [16] 
   and $\displaystyle\diamond$ are those of [11]. 
   }	
  \label{plasma_parameter}
\end{figure}

An understanding of dense plasmas is important to many areas of
physics such as astrophysics or solid state physics. Because 
a complete analytic theory has not yet been 
advanced, several simulation techniques have been developed
\cite{Ha73,CA80,HN93,PC94,KTR94} in order to determine properties
of plasmas. In such systems, however, the fermionic character and the 
long-range Coulomb forces cause many difficulties.\\
\indent
Quantum molecular dynamics (QMD) has been used to solve these
problems \cite{F90,FBS95}. This method has been developed as an extension
of classical molecular dynamics \cite{AT87}. In QMD, wave packets are 
used as an approximation of the wave function of particles.
Their dynamics can be derived from a time-dependent 
variational principle \cite{F90,RRB74}. It is possible to consider
bound states, exchange effects and correlations in this method.\\
\indent
The traditional QMD technique is based on Gaussian wave packets. The 
first simulations of this kind were carried out by Heller \cite{He75}. 
This technique was further developed in nuclear physics to
study scattering processes of nuclei \cite{F90,FBS95} and recently 
has been successfully applied to plasma physics by Klakow at el.
\cite{KTR94,Klakow_diss}. Both equilibrium
and non-equilibrium properties of dense plasmas can be studied.\\
\indent
In this paper we describe a new 
QMD model for partially ionized plasmas, 
which extends the previous work by incorporating 
free and bound states and transitions of the electrons.
In plasmas, such transitions can be caused by radiation and 
by collisions. Only
the latter transitions are important in the range of temperature and 
density that we are considering. The description of such 
processes is very
complicated, especially because the  microscopic treatment of the 
3-particle recombination is still an open question.\\
\indent
There have been recent attempts made to introduce transitions in QMD. 
In these it has been necessary to 
upgrade to Hamiltonian dynamics by use of additional elements. 
Ohnishi and 
Randrup have proposed to allow transitions on the basis of a stochastic 
dynamics \cite{OR94}. Tully has developed a branching concept in order
to study electronic transitions at surfaces \cite{Tu90}. In his approach 
the Hamiltonian dynamics was interrupted at certain points and 
a quantum mechanical transition was carried out.\\
\indent
In our model we start with a similar idea. In order to 
describe ionization
and recombination processes we implement two possible transitions:
A free electron can ionize an atom in a collision, and an ion and an
electron can recombine in a 3-particle collision of two free electrons
and an ion. We use the well-known cross section for the ionization 
process \cite{DE77,Lo67}. In order to model the rather complicated 
recombination process we have developed a preliminary model. We study 
the partially ionized, weakly degenerate
(n$_e\Lambda^3<$0.15) hydrogen
plasma at a density of 1.35$\times$10$^{22}$cm$^{-3}$  in the 
temperature range from 10000K to 200000K 
(see fig. \ref{plasma_parameter}). This is a much lower density than
those studied by Pierleoni et al. \cite{PM95} and Klakow et al. 
\cite{Klakow_diss}. Simulations based on 
this model converge
into an equilibrium of ionization and recombination 
with the degree of ionization in good agreement with theoretical
predictions.
\section{Basic assumptions}
\subsection{Quantum molecular dynamics}
Quantum molecular dynamics is a computational technique to study
quantum many body systems. It is based on an approximation for the
single particle wave function. Such a trail state $\Psi_{ {q_\nu} (t) }$
contains the essential degrees of freedom and the parameters
${q_\nu} (t)$ represent the coordinates in a generalized phase space.
The time evolution of the trail wave function is specified by the
dynamics of the time-dependent parameters ${q_\nu} (t)$.
For a chosen parameterization of the trial state, the equations of 
motion for the parameters can be derived from a 
time-dependent variational principle \cite{F90}:
\BE
\delta \int_{t_1}^{t_2} dt 
\left\langle \Psi_{q_\nu(t)}
\left| i \frac{d}{dt} 
- \hat{H} \right| 
\Psi_{q_\nu(t)} \right\rangle
\label{variational_principle}
= 0\quad.
\EE
The equations of motion can be written in the general form 
\cite{F90},
\begin{eqnarray}
{\dot{q}_\mu} = 
\sum_{\nu} A_{\mu\nu}^{-1} \frac{\partial H}{\partial {q_\nu}} \quad,
\label{Bewegungsgleichung}
\EE
in which the Hamilton function $H$ is defined as the expectation value
of the Hamiltionian $\hat{H}$,
\BE
  H\left({q_\nu}(t) \right) &=& 
  \left\langle \Psi_{q_\nu(t)} \left|
  \hat{H}
  \right|\Psi_{q_\nu(t)} \right\rangle \quad,
\EE
and where 
\BE
A_{\nu\mu} = 2 \; \mbox{Im}
\left\langle
\frac{\partial}{\partial q_\mu}
\Psi_{q_\mu(t)}
\left|
\frac{\partial}{\partial q_\nu}
\Psi_{q_\nu(t)}
\right.
\right\rangle
\label{Expl_Matrix}
\end{eqnarray}
is a skew-symmetric matrix \cite{F90,Mi96}. In certain trial wave 
functions, the parameters can be grouped into pairs of canonical 
variables and the matrix assumes the canonical form
\begin{equation}
A_{\nu\mu}^{-1} =
\left( \begin{array}{cc} 0 & -1 \\ 1 & 0 \end{array} \right)
\end{equation}
which leads to a Hamiltonian dynamics
for the parameters. Gaussian wave packets (GWP) obey this type of dynamics.

\subsection{Free electrons}

GWP have a long tradition in QMD 
\cite{F90,KTR94,He75,Klakow_diss,Li86,FBS95}. We use them
as trial wave function to describe the free electrons in the plasma:
\begin{equation}
\Psi_{GWP}(\vec{x})=\left(\frac{2}{\gamma^2 \pi}\right)^{3/4} 
\exp\left\{-\frac{\left(\vec{x}-\vec{r}\right)^2}{\gamma^2} 
+ i\vec{p}(\vec{x}-\vec{r})
\right\}\quad.
\label{GWP_eqn}
\end{equation}
The position $\vec{r}$ and the momentum $\vec{p}$ stand for the 
expectation values of the relevant quantum mechanical operators.
The parameter $\gamma$ determines the width of the wave packet.
Klakow et al. \cite{KTR94} made the first plasma simulation with 
GWP. In their model $\gamma$ is a complex parameter of the dynamics
and an additional degree of freedom of the trial wave function. We
have shown that this ansatz leads at high temperatures to a heat 
capacity per electron
 greater than $\frac{3}{2}k_B$ \cite{Mi96}.\\
\indent
Because of this principle problem we fixed the width of the GWP in our 
simulations. A condition for $\gamma$ can be derived from the 
Kelbg potential \cite{Ke63}:
\bearr
V^{Kelbg}(r)&=& \frac{{\bf e^2}}{r} 
\left[ 1-e^{-r^2/\lambda^2}+\frac{\sqrt{\pi} r}{\lambda}
  \left( 1-\erf \left\{ 
      {\textstyle \frac{r}{\lambda}}
 \right\} \right)
\right]\\
\nonumber
\lambda&=&\frac{\hbar}{\sqrt{m k_B T}}
\quad\mbox{valid for}\quad
\xi = \frac{\bf e^2}{\hbar}\sqrt{\frac{2m}{k_B T}} \ll 1,
\eearr
which should approximately coincide with the interaction potential 
of two GWP \cite{Klakow_diss} 
\begin{eqnarray}
  V_{ee,GWP-GWP}(r) &=&  
  \frac{{\bf e^2}}{r}
  \erf\left\{ \frac{r}{\gamma} \right\}\quad,
  \label{V_GG}
\end{eqnarray}
Assuming 
$\LL.V_{ee,GWP-GWP}\RR|_{r=0}=\LL.V^{Kelbg}\RR|_{r=0}$
we fitted 
\BE
\gamma = \frac{2}{\pi}\lambda\qquad.
\label{gamma_fit}
\EE
This ansatz leads to canonical equations of motion for the parameters
$\vec{r}$ and $\vec{p}$.
\subsection{Bound electrons}
In our simulation we consider the 1s-ground state of hydrogen
explicitly. The best approximation for the ground state with GWP 
leads to an energy of 11.5 eV instead of 13.6 eV
\cite{Klakow_diss}. GWP with a dynamics derived from 
(\ref{variational_principle}) have a continuous spectrum of 
excitation and a
 gap is missing. This results into unphysical excitations in many
particle simulations and strongly influences the 
thermodynamic properties of the system \cite{Mi96}.
That is why we introduce a 1s wave function (1sWF) to describe bound states,
\BE
\Psi_{1sWF}(\vec{x})&=& \frac{1}{\sqrt{\pi a_0^3}}
\exp\left\{ 
-\frac{1}{a_0} \left| \vec{x}-\vec{r}_{I} \right|
\right\}
\label{1sWF_eqn}
\EE
where $\vec{r}_{I}$ is the position of the core. This wave function
is an eigenstate of the Schr\"odinger equation.
\subsection{Simulation of Hydrogen Plasma}
The hydrogen plasma consists of ions with positive charge and electrons
with negative charge. In order to incorporate quantum mechanical
effects, the electrons in our model are treated as wave packets. The ions,
which have much greater mass, are considered to be classical particles. 
We distinguish between free and bound states 
of the electrons by representing free electrons by GWP (\ref{GWP_eqn}) and
bound electrons by 1sWF (\ref{1sWF_eqn}). 
Our method is based on the Hamilton function $H$, which consists of the 
several parts,
\begin{equation}
H = T_i + T_e + V_{ii} + V_{ee} + V_{ei}\quad.
\end{equation}
\noindent
The classical ions contribute kinetic and potential energy,
\begin{eqnarray}
  T_i &=& \sum_{k} \frac{\vec{p}_k^{\;2}}{2M_k}\quad,
  \\
  V_{ii} &=& \sum_{k<l} \frac{{\bf e^2}}
  {\left| \vec{r}_k-\vec{r}_l \right|} + 
  \Phi_{Ewald}\left(\left| \vec{r}_k-\vec{r}_l \right|\right)\quad,
\end{eqnarray}
where the Ewald potential is used for
periodic boundary conditions \cite{Ha73,AT87}.
The following terms depend on the trial wave functions of the electrons,
\begin{eqnarray}
T_e &=& \sum_{k} T_{e,k}\quad,
\\
V_{ee} &=& \sum_{k<l} V_{ee,k,l}+ 
\Phi_{Ewald}\left(\left| \vec{r}_k-\vec{r}_l \right|\right)\quad,
\\
V_{ei} &=& \sum_{k,l} V_{ei,k,l}- 
\Phi_{Ewald}\left(\left| \vec{r}_k-\vec{r}_l \right|\right)\quad,
\\
T_{e,k} &=& 
\left< \Psi_k \left|  \frac{\vec{\hat{p}}_k^{\;2}}{2m} 
\right| \Psi_k\right>\quad,
\\
V_{ee,k,l} &=& 
\left< \Psi_k(\vec{x}) \Psi_l(\vec{y}) \left|
\frac{{\bf e^2}}{ \left| \vec{x}-\vec{y} \right| }
\right| \Psi_l(\vec{y}) \Psi_k(\vec{x}) \right>\quad,
\label{V_ee}
\\
V_{ei,k,l} &=& 
\left< \Psi_k(\vec{x}) \left| 
\frac{-{\bf e^2}}{ \left| \vec{x}-\vec{r}_l \right| }
\right| \Psi_k(\vec{x}) \right>\quad.
\label{V_EI}
\end{eqnarray}
For GWP and 1sWF, this leads to
\begin{eqnarray}
T_{e,GWP} &=&
\frac{1}{2m}\left(\frac{3}{\gamma^2} 
+ {\vec{p}}^{\;2}\right)\quad,
\label{E_C_GWP}
\\ 
T_{e,1sWF} &=& 
\frac{1}{2m} \left(
\frac{1}{a_0^2} + {\vec{p}}^{\;2}\right)\quad,
\\
V_{ei,GWP} &=&  
-\frac{{\bf e^2}}{r}
\erf\left\{ \frac{\sqrt{2}r}{\gamma} \right\}\quad,
\label{GWP_V_EI}
\\
V_{ei,1sWF} &=&  
-\frac{{\bf e^2}}{r}
\left[1-
e^{ -2 r /a_0}
\left( \frac{r}{a_0}   + 1 \right)
\right]\quad,
\end{eqnarray}
\noindent
where $r$ is the distance between the particles.
Different wave functions lead to different potentials for the
electron-electron interaction. The interaction of two GWP is given by 
(\ref{V_GG}). The interaction of a 1sWF with either a GWP or a second 1sWF
is given by 
\begin{eqnarray}
  V_{ee,GWP-1sWF} &=& 
  \frac{{\bf e^2}}{r} \erf \left\{ r/d \right\} - 
  {\bf e^2}\frac{e^{-r^2/d^2}}{2 r }
  \left[ 
    f\left( \frac{d}{a_0} + \frac{r}{d}\right) - 
    f\left( \frac{d}{a_0} - \frac{r}{d}\right)
  \right] \quad,
  \label{V_GH}
\\
  \nonumber
  & &
  f(x) = e^{x^2} \mbox{erfc}\{x\} 
  \left( x\frac{d}{a_0} -1 \right)
  \hspace{0.25cm}
  ,
  \hspace{0.25cm}
  d = \frac{\gamma}{\sqrt{2}}\quad,
\\
  V_{ee,1sWF-1sWF} &=& 
  \frac{{\bf e^2}}{a_0 \rho} e^{-2\rho}\LL[ 
  1 + \frac{5}{8}\rho - \frac{3}{4}\rho^2  - \frac{1}{6}\rho^3 
  \RR]
  \hspace{0.25cm}
  ,
  \hspace{0.25cm}
  \rho = \frac{r}{a_0}\quad.
  \label{V_HH}
\end{eqnarray}
An atom in the simulation consists of a proton and an electron in the 
1s-ground state. The interaction potentials of atoms are obtained
by adding the contributions from both parts.\\
\indent
The ansatz (\ref{GWP_eqn}) reveals the extra term 3/(2m$\gamma^2$) 
in the kinetic energy of a GWP (\ref{E_C_GWP}), which has its
origin in the shape of the GWP. Since $T_{e,GWP}$ is not consistent with
the classical limit of free particles for high temperatures 
$\vec{p}^{\;2}/2m$ we omit this term.\\
\section{Thermodynamic Model}
\begin{figure}[htb]
  \vspace*{-1.5cm}
  \centerline{\psfig{figure=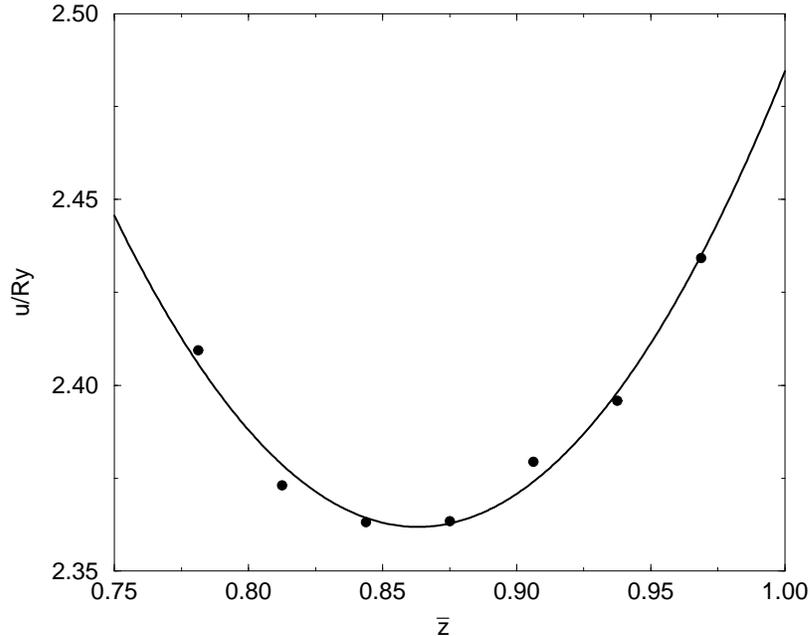,width=12cm,angle=270}}
  \vspace*{-0.5cm}
  \caption{ 
    Internal energy $u$ as function of the degree of ionization
    at constant entropy.
  }
  \label{u_s_min}
\end{figure}

In this paragraph it will be shown how the QMD can be used to determine
the degree of ionization $\bar{z}$ (the average charge number)
of the plasma. In the usual approach 
one derives the degree of ionization from the free energy $F$, 
which is minimal at constant 
volume and temperature. Alternatively it can be determined by using the
condition that the internal energy $U$ is minimal at constant entropy and 
volume.
\BE
   \LL( \frac{\partial F}{\partial \bar{z}} \RR)_{T,V,n} = 0
  \qquad,\qquad
  \LL( \frac{\partial U}{\partial \bar{z}} \RR)_{S,V,n} = 0\quad.
\EE
The free energy is not available in QMD simulations. That is why the 
second condition is used to find the equilibrium-degree of ionization.
In order to find the minimum of the internal energy $U$ 
a number of separate simulations with different degrees of ionization
were carried out without including transitions between free
and bound states.\\
\indent
The entropy must be the same in all of these simulations. This 
requires a different temperature in every simulation. It
can be obtained from entropy function $S$, which has to be known 
in our method. We use the entropy of the ideal hydrogen plasma,
\bearr
  S(N_e,N_i,N_0) = \sum_k N_k k_B \LL[ 
    \frac{5}{2} - \log \LL(\frac{\Lambda_k^3 N_k}{g_k V} \RR) 
  \RR]
  \quad,\quad
  \Lambda_k = \frac{h}{(2\pi m_k k_B T )^{1/2}}
  \label{S_id}
  \quad.
\eearr
$N_k$ is the number of particle in the volume $V$ and $g_k$ is the degeneracy
parameter due to the spin, where the index $k$ runs over all particle types. 
If one describes a hydrogen plasma in the chemical picture 
\cite{rotesbuch} one has to consider atoms ($g_0$=2), ions ($g_i$=1)
and free electrons ($g_e$=2). The degree of ionization is defined by 
$\bar{z}=N_e/N$, where $N_e$=$N_i$ and $N=N_0+N_i$.
\\
\indent
Under the condition of fixed entropy the appropriate temperature 
can be derived by equation (\ref{S_id}) for any degree of ionization.
In the simulations we use a velocity-scaling procedure 
\cite{AT87}, which forces the system towards the desired temperature. 
Then the internal energy is obtained by averaging 
over many phase space configurations of the system. 
In this way the internal energy can be determined point by 
point on an adiabatic and its
minimum can be found. The exact value for the minimum is obtained by
quadratic approximation (fig. \ref{u_s_min}). 
The temperature belonging to this minimum 
has to be calculated from (\ref{S_id}) since the temperature in all 
simulations is different.
\subsection*{Results}
We study a partially ionized hydrogen plasma at a density of 
1.35$\times$10$^{22}$cm$^{-3}$ in the temperature range from 10000K
to 200000K. The plasma is weakly degenerate: $n_e\Lambda^3 < 0.15$,
where $n_e$ is the density of the free electrons derived from
the ideal Saha equation \cite{rotesbuch}. We consider 
32 electrons and 32 ions in our simulations.\\
\begin{figure}[hbt]
  \vspace*{-1.5cm}
  \centerline{\psfig{figure=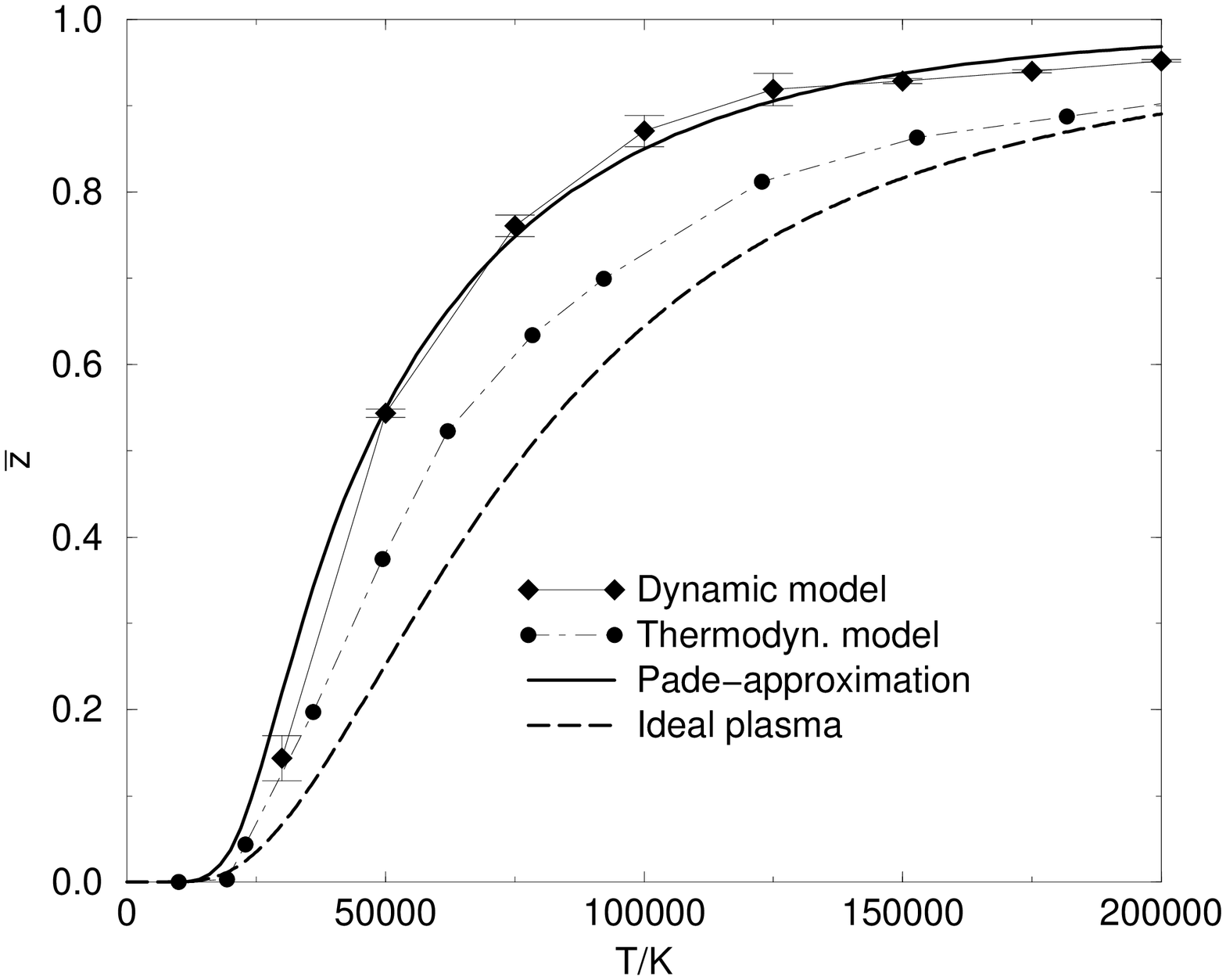,width=12cm,angle=270}}
  \vspace*{-0.5cm}
  \caption{Degree of ionization versus temperature 
  }
  \label{fig_alpha}
\end{figure}
\indent
In figure \ref{fig_alpha}, the degree of ionization from the 
simulation in the thermodynamical model is compared with results from
the Pad\'e approximation in the chemical picture (PACH) \cite{Binz} and
from the ideal Saha equation. The PACH approach is based on Pad\'e formulae 
for the thermodynamic functions \cite{ER85b} and on the non-ideal 
Saha equation \cite{rotesbuch}. All methods show no ionization 
below 20000K,
then the degree of ionization increases sharply. The values from
our simulation and from the PACH are higher than 
that of the ideal plasma because the Coulomb interaction leads to
a reduction of the effective ionization energy. At higher temperatures,
the Coulomb interaction is less important, therefore the difference
between the results from our simulation and the ideal Saha 
equation becomes smaller. 
The values from the 
PACH are still 6\% higher because the 
Planck-Larkin partition function, which takes into account excited 
states is used in these formulas 
\cite{rotesbuch}.\\
\begin{figure}[p]
  \vspace*{-1.5cm}
  \centerline{\psfig{figure=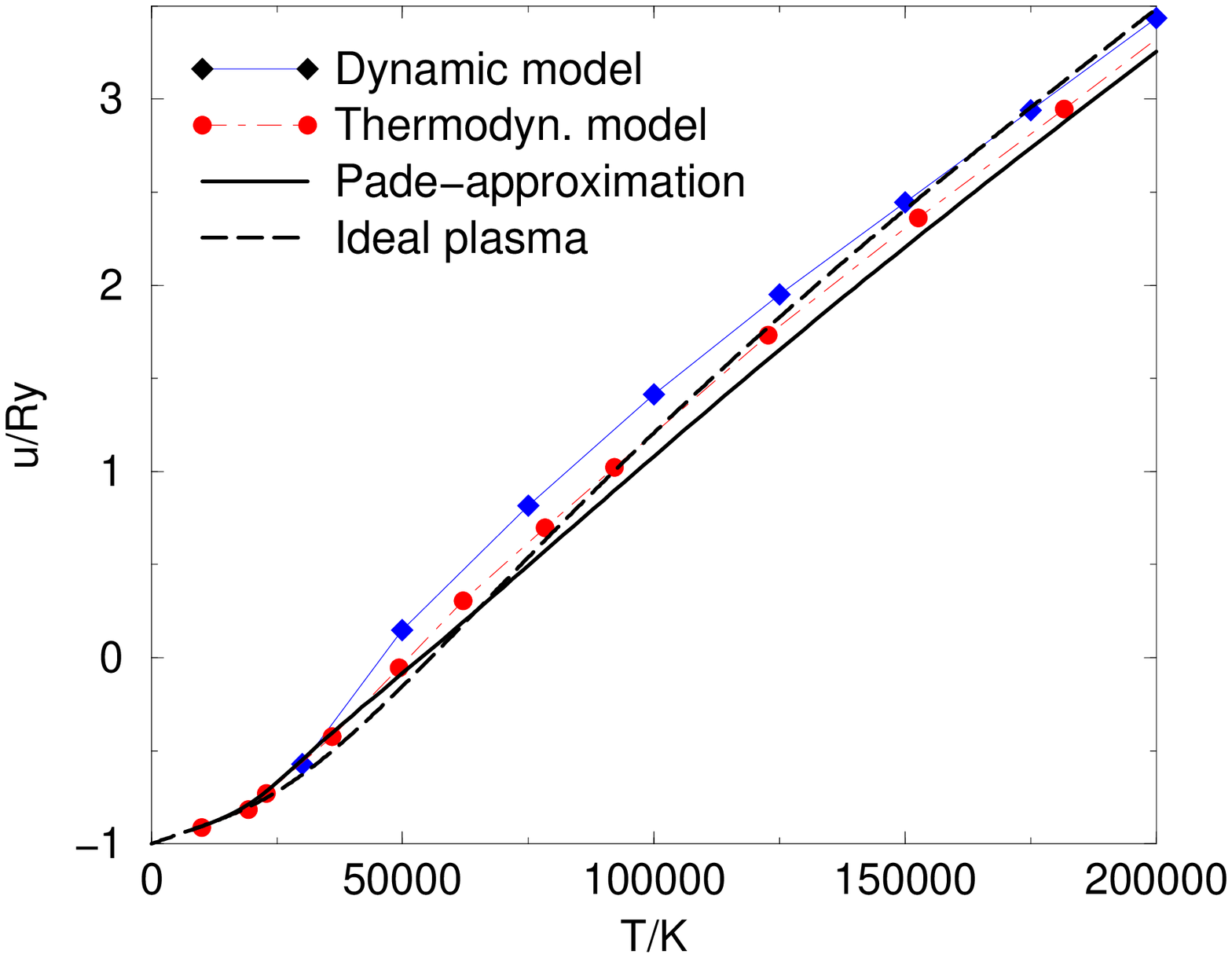,width=12cm,angle=0}}
  \vspace*{-1cm}	
  \caption{Internal energy u versus temperature
  }
  \label{fig_E}
%
  \centerline{\psfig{figure=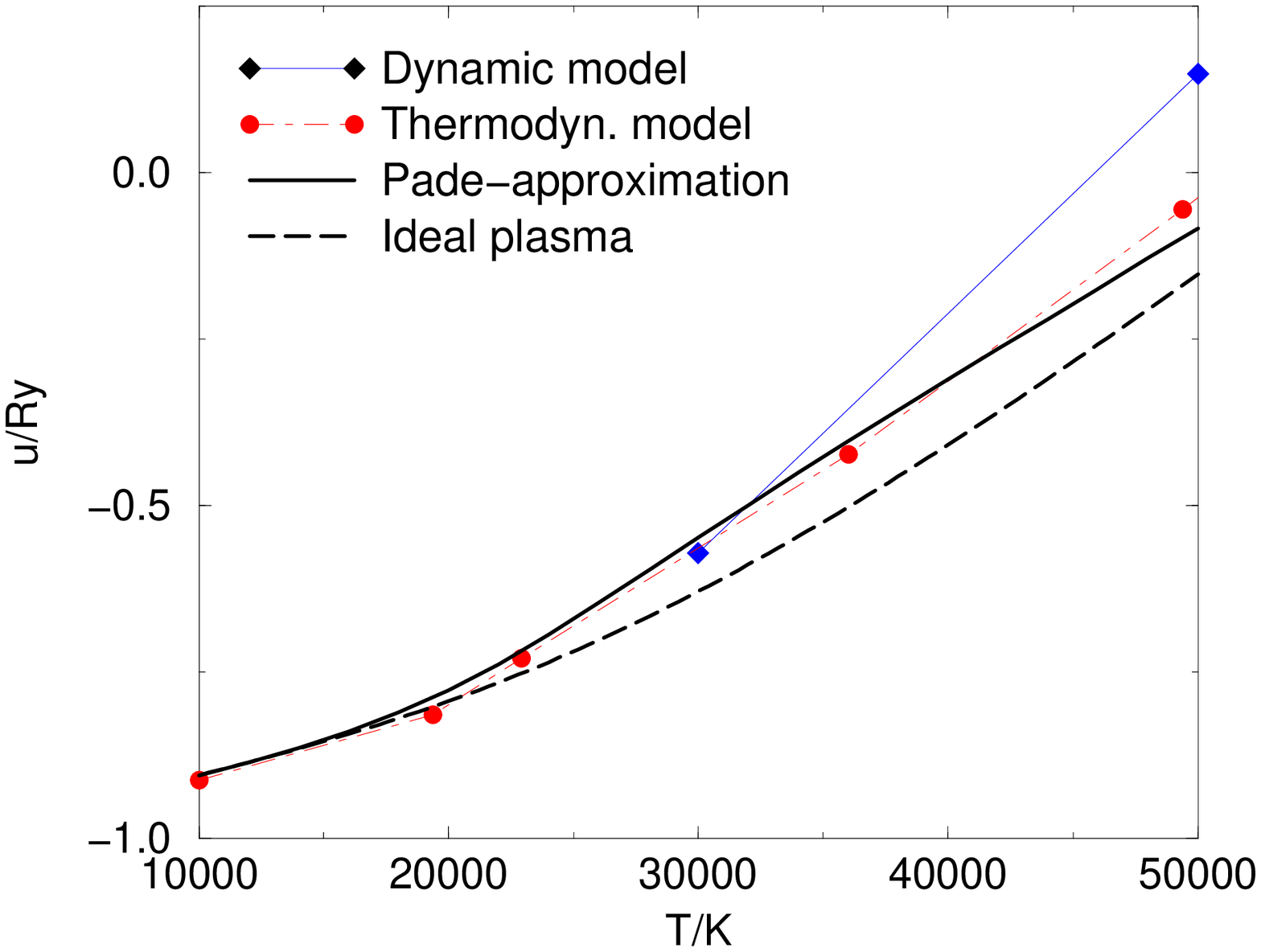,width=12cm,angle=0}}
  \vspace*{-1cm}	
  \caption{Internal energy u versus temperature
}
  \label{fig_E2}
\end{figure}
\indent
The internal energy per particle from the simulation in the thermodynamic
model $u_{th}$, from the
PACH $u_{PACH}$ and from the ideal plasma $u_{id}$ 
are shown in figure
\ref{fig_E} and \ref{fig_E2}. Under 20000K hydrogen
behaves like an atomic gas, so $u$ increases like 3/2~k$_B$T. 
All three theories coincide in this region.
With increasing temperature $u_{th}$ and
$u_{PACH}$ rise more quickly than $u_{id}$ because of the higher degree of
ionization, which leads to addtitional contributions to the kinetic energy. 
At high temperatures, negative contributions from
the Coulomb interaction determine the corrections to $u_{id}$. So 
$u_{th}$ and $u_{PACH}$ intersect with $u_{id}$ at an
intermediate temperature.\\
\indent
The agreement between the 
$u_{th}$ and $u_{PACH}$  is fairly good in the region 
10000K\leqs T\leqs 50000K. For higher temperatures, deviations start to occur. 
Above 75000K, $u_{th}$ 
and $u_{PACH}$ show the same behavior but differ by a constant. 
The PACH predicts a smaller energy although 
the degree of ionization is higher. One possible explanation is that there
are no interactions of charged and neutral particles taken into account
in these formulas. Further we have to mention that our approach 
neglects excited states, which are taken into account in the 
PACH approach \cite{Binz}.

\section{Dynamic Equilibrium of Ionization and Recombination}

There have been several attempts to model reactions by QMD. 
Tully developed a branching concept for transitions between
different electronic states \cite{Tu90}. In this approach, the transition 
probabilities were
derived from a Greens function technique. Hamiltonian dynamics is still used
most of the simulation. Only in the case of a transition, dynamics are 
interrupted, new initial conditions for the particles are formulated and then 
the dynamics are continued.\\
\indent
In our simulation we must deal with ionization and recombination 
processes. If an electron
with an impact energy greater than 1 Ry collides with an atom, ionization
can take place. In order to include such transitions in QMD, knowledge
of the microscopic dynamics of such a process is necessary. The 
ionization of one hydrogen atom in a plasma was studied in \cite{EFP95}.
Here we investigate the much more complicated case where many atoms are 
present and include ionization as well as recombination.
The ionization cross section for hydrogen is known various theoretical
and experimental predictions \cite{Lo67,DE77}. We use the 
semi-empirical formula \cite{Lo67,EFP95}:
\BE
\sigma(E) = c\frac{\ln\LL(\frac{E}{I}\RR)}{E I}
\LL[1-\exp\LL(1-{\textstyle\frac{E}{I}}\RR)\RR]
\quad\mbox{if}\quad E>I
\;,\; I = 1\:Ry\;,\; c=7.7~Ry^2a_0^2\;.
\label{eqnSigma}
\EE
\indent
Adapting the cross section to the wave packet model we imagine
a sphere of radius $\sqrt{\sigma_I(E)/\pi}$ around every atom. If
a free electron enters this sphere with an impact energy $E$ greater 
the 1 Ry the ionization of the atom occurs. The electron in the 1s 
ground state is replaced by free GWP, which is place at the opposite point
on the sphere. The momenta are chosen arbitrarily under conservation
of energy.\\
\indent
A microscopic description of the recombination process is more difficult.
However special information on recombination is available from various
investigations. Besides the cross sections in different approximations
which were derived from scattering theory, also the global recombination
rate $\beta$ is known from the theory of the rate coefficients 
\cite{DE77,LE93}. 
A microscopic description of the recombination process in QMD has
not been derived yet. As a 
preliminary model we propose the following mechanism: If two electrons
are simultaneously in a sphere of radius $\rho_R$ = 1 a$_B$ around
an ion the recombination will take place. One electron is moved into
the ground state and the other changes its momentum so that the total
energy is conserved.

\subsection*{Results}
We carried out simulations the the same density from 30000K (128
ions and 128 electrons) to 200000K (32 ions and 32 electrons). The system
converges into a dynamic equilibrium of ionization and recombination. 
The degree of ionization we get from this simple model 
is surprisingly good (fig.\ref{fig_alpha}). 
It agrees well with the PACH approach in the considered temperature range. 
We do not claim that we give with our simple microscopic model a rigorous 
description of the recombination process but, rather, we deduce that our 
mechanism leads to a correct global rate $\beta$ in the temperature
range studied here. Correct in context means that the average
number of recombination processes per time unit is in agreement with the 
number ionization processes so that the detailed balance is satisfied.

In our model, we have found a 
dependence of $\beta$ on 
$\rho_R$ and a setting of $\rho_R = 1 a_0$ turned out to be a good
estimate.\\
\indent
The internal energy $u$ derived from this simulations leads to
qualitatively correct results (fig.\ref{fig_E}) but quantitatively 
the results only agree well up to temperature of 40000K (fig.\ref{fig_E2}). 
Beyond of that our results are systematically high in comparison to the PACH,
which may be caused by an insufficient description of the
electron-ion interaction.  because our results were generally too high
in comparison to the PACH model. Evidently simulations above 40000K require
a more precise description of the interaction between free
electrons and ions than that given in our QMD model. 
\section{Conclusion}
We expect that the model presented in this paper will prove to be 
a promising 
approach for the description of plasmas in the partially ionized region. 
We have substantially extended
the traditional QMD model 
with the consideration of free and 
bound states. We have shown how thermodynamical functions can be derived
from a model with free and bound states but still without transitions.
Furthermore we presented preliminary model for the dynamic description 
of ionization and recombination in plasmas.
As first test we calculated the degree of ionization and the internal energy
of the hydrogen plasma.\\
\indent
The model can be improved in several ways. First of all a detailed
quantum mechanical description of the 3-body recombination process
would be useful in future studies. Furthermore it would be interesting
to study the formation of molecules \cite{Mi96}. One must also
take into account the fermionic character of the electron
wave function \cite{FBS95,Klakow_diss}.
Our results offer hope that it soon will be possible to describe
hydrogen in the whole temperature range from the molecular 
gas to the completely ionized plasma.
\section{Acknowledgments}
We acknowledge useful and stimulating discussions with J. Ortner,
A. F\"orster, and D. Beule.\\
\newpage
\bibliographystyle{unsrt}

\end{document}